\documentclass[journal]{IEEEtran} 
\IEEEoverridecommandlockouts
\usepackage{cite}
\usepackage{amsmath,amssymb,amsfonts}
\usepackage{algorithm,algpseudocode}
\usepackage{graphicx}
\usepackage{textcomp}
\usepackage{xcolor}
\usepackage{graphicx}
\usepackage{titlesec}
\titlespacing\section{0pt}{0pt plus 4pt minus 2pt}{0pt plus 2pt minus 2pt}
\titlespacing\subsection{0pt}{1pt plus 4pt minus 2pt}{1pt plus 2pt minus 2pt}

\usepackage{subcaption}
\def\BibTeX{{\rm B\kern-.05em{\sc i\kern-.025em b}\kern-.08em
    T\kern-.1667em\lower.7ex\hbox{E}\kern-.125emX}}
\begin{document}

\title{V2V Communications Using Blockchain-Enabled 6G Technology and Federated Learning}

\author{
    \IEEEauthorblockN{Tahir H. Ahmed\IEEEauthorrefmark{2}, Jun Jiat Tiang\IEEEauthorrefmark{2}, Azwan Mahmud\IEEEauthorrefmark{2}, Dinh-Thuan Do\IEEEauthorrefmark{3}, Truong Tran\IEEEauthorrefmark{4}, Shahid Mumtaz\IEEEauthorrefmark{5}}\\
    \IEEEauthorblockA{\IEEEauthorrefmark{2}\textit{Centre of Wireless Technology, Multimedia University, Malaysia.}}\\
    \IEEEauthorblockA{\IEEEauthorrefmark{3}\textit{School of Engineering, University of Mount Union, USA.}}\\
    \IEEEauthorblockA{\IEEEauthorrefmark{4}\textit{School of Science, Engineering, and Technology, Penn State University, USA.}}\\
    \IEEEauthorblockA{\IEEEauthorrefmark{5}\textit{Department of Applied Informatics, Silesian University of Technology, Akademicka 16, 44-100 Gliwice, Poland, and Nottingham Trent University, Department of Engineering.}\\
    Emails: \IEEEauthorrefmark{2}\{1211405181@student., jjtiang@, azwan.mahmud@\}mmu.edu.my, \IEEEauthorrefmark{3}doth@mountunion.edu,\IEEEauthorrefmark{4}truong.tran@psu.edu, \IEEEauthorrefmark{5} dr.shahid.mumtaz@ieee.org}}

\maketitle

\begin{abstract}
      This study proposes an interesting approach for vehicle-to-vehicle (V2V) communication, which integrates blockchain technology, federated learning (FL), and allocation optimization of latency and resources. The research evaluates the proposed system using various performance metrics such as packet delivery ratio (PDR), model accuracy, and latency and demonstrates its superiority over existing techniques. Furthermore, the system provides enhanced security through consensus optimization and k-anonymity for data privacy. Overall, the proposed system is a promising solution for efficient and secure V2V communication in the era of connected and autonomous vehicles. Moreover, the proposed approach achieves higher reliability, lower latency, and better resource utilization compared to traditional 5G. 
\end{abstract}

\begin{IEEEkeywords}
    V2V, 5G/6G, Beyond 5G, Federated Learning, Blockchain
\end{IEEEkeywords}

\section{Introduction \& Related Work}
The evolution of automotive networking combined with advanced wireless communication methods has spurred the creation of Vehicle-to-Vehicle (V2V) communication systems that enhance road safety, optimize traffic flow, and minimize environmental impact. As the deployment of sixth-generation (6G) wireless networks draws nearer, it becomes imperative to integrate blockchain and FL technologies to ensure secure and efficient V2V communication systems \cite{shi2022drl}. Previous research has delved into various aspects of V2V communication. For instance, \cite{b1} studied an FL method to accelerate training in V2V systems, while \cite{b2} proposed the FLOR framework that can offload and manage resources with low latency across multiple networks. Additionally, \cite{b3} focused on optimizing transmission resource allocation and mode selection using a deep reinforcement learning-based decentralized algorithm.

However, several gaps remain in the literature. Works like \cite{9741761} emphasized task computation delay minimization but missed out on the potential benefits of federated learning. Similarly, while \cite{9686954} highlighted resource allocation in IoT networks, the integration of blockchain for security wasn't considered. The approach by \cite{9749770} offered advancements in error detection for V2X data but did not leverage the strengths of federated learning for real-time corrections. Moreover, the DRL approach for 6G V2V communications in \cite{electronics12102294} did not factor in the significance of both federated learning and blockchain.

In contrast to existing research, our study seeks to bridge these gaps by holistically incorporating blockchain and FL. Conventional studies often bypass the combined benefits of these technologies, especially in terms of security, data privacy, and resource optimization \cite{b1, b2, b3}. We provide a decentralized, tamper-proof, and collaborative platform for V2V communication, leveraging these technologies to pave the way for intelligent transportation systems \cite{mollahasani2021dynamic, 9798266, 9990577, cui2021secure}.

Our contributions include:
\begin{itemize}
    \item We develop a two-stage optimization approach to enhance the performance of blockchain-assisted V2V communication networks by minimizing total latency and ensuring minimum SINR at each node. To this end, we formulate a mixed-integer nonlinear programming (MINLP) problem that addresses both communication channel allocation and transmission power control for each vehicle. The MINLP problem is solved using a two-stage approach consisting of Nonlinear Programming Relaxation, Integer recovery, and Link Selection, effectively optimizing V2V communication within the FL framework.
    \item A novel mathematical model for FL in 6G V2V networks is proposed, addressing the trade-off between local data privacy and global model accuracy. This approach trains local models using spatio-temporal-aware adaptive learning rates, determined by SINR and vehicle distance. Consensus optimization synchronizes local models among vehicles while preserving privacy, with consensus weights based on SINR and vehicle distance.
    \item We propose a mathematical model and a two-stage optimization algorithm for THz 6G networks with blockchain-assisted FL in V2V communication. The algorithm combines Multi-Objective Particle Swarm Optimization (MOPSO) for FL and resource allocation and Blockchain-Assisted Consensus Optimization (BACO) for optimizing the blockchain consensus process. The framework integrates blockchain, FL, latency, and resource allocation models, leading to efficient consensus, minimized latency, and maximized resource utilization in V2V communications.
\end{itemize}
The paper's subsequent sections present our approach, evaluate performance metrics, and conclude the study.

\section{Methodology}\label{M}

This section describes our methodology  of Blockchain-Assisted 6G System and FL to handle the V2V Communications.

Let $G(V, E)$ represent the V2V communication graph consisting of vertices (nodes) set $V$ and communication links set $E$ (edges.) The nodes set $V$ is combined of two subsets: $V_v$ for vehicles and $V_r$ for Roadside units (RSUs), where $|V_v| = N$ and $|V_r| = M$. The connectivity of the graph is represented by the adjacency matrix $A \in \mathbb{R}^{N \times N}$, where $a_{ij} = 1$ if vehicles $i$ and $j$ are connected, and $a_{ij} = 0$ otherwise.

Let $C$ be the number of communication channels available in the network. Each communication channel $c \in {1, \dots, C}$ has a bandwidth $B_c$ and a maximum transmission power $P_{\text{max},c}$.

Let $d_{ij}$ represent the distance between nodes $i$ and $j$. The path loss model for the channel gain $h_{ij,c}$ between nodes $i$ and $j$ on channel $c$ is given by:
\begin{equation}
\text{PL}(d_{ij}) = {PL_0} + 10 * \alpha * \log{10}(d_{ij} / d_0),
\end{equation}
where $PL_0$ is the path loss at a reference distance $d_0$, and $\alpha$ is the path loss exponent. The channel gain $h_{ij,c}$ is related to the path loss by:
\begin{equation}
h_{ij,c} = 10^{-\text{PL}(d_{ij}) / 10}.
\end{equation}
The signal to interference plus noise ratio (SINR) at node $i$ on channel $c$ is given by:
\begin{equation}
\text{SINR}{i,c} = \frac{P{\text{tx},c} * h_{ij,c}}{I_{i,c} + N_0},
\end{equation}
where $P_{\text{tx},c}$ is the transmission power of node $j$ on channel $c$, $I_{i,c}$ is the interference at node $i$ on channel $c$, and $N_0$ is the noise power.
The interference $I_{i,c}$ is the sum of interference from all other nodes transmitting on channel $c$:
\begin{equation}
I_{i,c} = \sum_{k \in V, k \neq j} P_{\text{tx},k} * h_{ik,c}.
\end{equation}
The data rate $R_{ij,c}$ between nodes $i$ and $j$ on channel $c$ is given by the Shannon capacity formula:
\begin{equation}
R_{ij,c} = B_c * \log_2(1 + \text{SINR}_{i,c}).
\end{equation}

\subsection{Blockchain Latency Model}



This section describes the total latency optimization  related to V2V communication in a blockchain-assisted 6G network.
We consider the following sources of latency in the blockchain-assisted 6G network:

\begin{itemize}
\item Propagation delay $T_p$: Time taken for a signal to travel between nodes.
\item Transmission delay $T_t$: Time taken to transmit a block of size $S_b$.
\item Block validation delay $T_v$: Time taken to validate a block.
\end{itemize}

The total latency $T_{\text{total}}$ is given by:
\begin{equation}
T_{\text{total}} = T_p + T_t + T_v,
\end{equation}
where $T_p = \frac{d_{ij}}{v_p}$, with $v_p$ being the propagation speed, $T_t = \frac{S_b}{R_{ij,c}}$, $T_v$ is a function of the computation capabilities of the validating nodes.

The objective is to minimize the total latency $T_{\text{total}}$ while ensuring adequate SINR at each node. The optimization problem is formulated as:
\begin{align}
\text{minimize: } & T_{\text{total}} \nonumber \\
\text{subject to:} \nonumber \\
& \text{SINR}_{i,c} \geq \text{SINR}_{\text{min}}, \forall i \in V_v, \forall c \in \{1, \dots, C\} \label{eq:constraint1} \\
& 0 \leq P_{\text{tx},c} \leq P_{\text{max},c}, \forall c \in \{1, \dots, C\} \label{eq:constraint2} \\
& a_{ij} \in \{0, 1\}, \forall i, j \in V_v \label{eq:constraint3}
\end{align}
where $\text{SINR}_{\text{min}}$ is the minimum SINR threshold required for reliable communication. 

The allocation of communication channels and control of transmission power for each vehicle is essential \cite{9793590}. The total number of available channels is represented by $C$, and the transmission power of vehicle $i$ on channel $c$ at iteration $t$ is represented by $P_{i,c}^{(t)}$. Binary variables $a_{ij,c}^{(t)}$ are introduced to indicate whether vehicle $i$ communicates with vehicle $j$ on channel $c$ at iteration $t$. The optimization problem is a mixed-integer nonlinear programming (MINLP) problem because of the integer constraints on $a_{ij}$ and the nonlinear relationships between variables. We design the two-state approach below to this problem.

\textbf{Stage 1 - Nonlinear Programming Relaxation:}
In this stage, the integer constraint in Equation (\ref{eq:constraint3}) is initially relaxed to obtain a continuous constraint, as follows:
\begin{equation}
0 \leq a_{ij,c}^{(t)} \leq 1, \forall i, j \in {1, \dots, N}, \forall c \in {1, \dots, C}.
\end{equation}
By solving this nonlinear programming problem using gradient-based or derivative-free optimization algorithms, the latency in the integrated optimization problem can be effectively reduced. Examples of such algorithms include interior point methods, sequential quadratic programming, and particle swarm optimization.

\textbf{Stage 2 - Integer Recovery and Link Selection:}
Upon obtaining the solution to the relaxed problem, the integer values for $a_{ij,c}^{(t)}$ are recovered by applying a threshold.
\begin{equation}
a_{ij,c}^{(t)} =
\begin{cases}
1, & \text{if } a_{ij,c}^{(t)} \geq \theta \\
0, & \text{otherwise}
\end{cases},
\end{equation}
where $\theta \in [0, 1]$ is a threshold value that can be adjusted to control the trade-off between communication reliability and resource utilization.

Utilizing the recovered integer values, communication links are selected to minimize the total latency. The link selection process must ensure that each vehicle has a minimum of one communication link to participate in the blockchain-assisted 6G network. This two-stage approach addresses the total latency optimization problem in the context of V2V communication.



\subsection{FL Framework Incorporating 6G Technology}
In this section, we develop a novel mathematical model for a FL framework in the context of 6G V2V networks, considering the unique characteristics and requirements of 6G technology. We consider a network of $N$ vehicles that collaborate in training a global model while preserving their local data privacy.

\subsubsection{Local Model Training with Spatio-Temporal-aware Adaptive Learning Rates}

In the proposed framework, each vehicle $i$ trains its local model $w_i^{(t)}$ using its own dataset $D_i$, where $t$ is the iteration number. To exploit the spatio-temporal aspects of 6G technology, we propose adaptive learning rates based on the distance and communication quality among vehicles. The adaptive learning rate $\eta_i^{(t)}$ for vehicle $i$ at iteration $t$ is calculated as:
\begin{equation}
\eta_i^{(t)} = \eta_0 \frac{\text{SINR}{i}^{\beta}}{d{ij}^{\gamma}},
\end{equation}
where $\eta_0$ is the initial learning rate, $\text{SINR}{i}$ is the signal-to-interference-plus-noise ratio for vehicle $i$, $d{ij}$ is the distance between vehicles $i$ and $j$, and $\beta$ and $\gamma$ are positive tuning parameters that control the impact of communication quality and distance on the learning rate, respectively.

The local model update for vehicle $i$ at iteration $t$ is expressed as:
\begin{equation}
w_i^{(t+1)} = w_i^{(t)} - \eta_i^{(t)} \nabla F_i(w_i^{(t)}),
\end{equation}
where $\nabla F_i(w_i^{(t)})$ is the gradient of the local loss function for vehicle $i$.
\subsubsection{Consensus Optimization for Global Model Synchronization}
To synchronize the local models among vehicles while preserving data privacy, we propose a consensus optimization approach that leverages the 6G-enabled V2V communication capabilities. At each iteration $t$, vehicle $i$ communicates with its neighbors in the V2V network to exchange model information. The consensus optimization update for vehicle $i$ at iteration $t$ is expressed as:

\begin{equation}
w_i^{(t+1)} = w_i^{(t)} + \sum_{j \in \mathcal{N}i} \lambda{ij}(w_j^{(t)} - w_i^{(t)}),
\end{equation}
where $\mathcal{N}i$ is the set of neighboring vehicles of vehicle $i$, and $\lambda{ij}$ is a non-negative consensus weight between vehicles $i$ and $j$. The consensus weights are determined based on the communication quality and distance among vehicles, as:
\begin{equation}
\lambda_{ij} = \frac{\text{SINR}{ij}^{\beta'}}{d{ij}^{\gamma'}},
\end{equation}
where $\beta'$ and $\gamma'$ are positive tuning parameters that control the impact of communication quality and distance on the consensus weights, respectively.
\subsection{Latency and Resource Allocation Model for Terahertz 6G with Blockchain and FL}

In this section, we present the mathematical model for latency and resource allocation in terahertz (THz) 6G networks, incorporating blockchain-assisted FL for V2V communication. This model aims to minimize the latency while maximizing the resource utilization, considering the specific parameters and constraints from the blockchain and FL framework.

Let $B_i$ denote the available bandwidth for vehicle $i$ in the THz 6G network, $T_{p_i}$ be the processing time required for vehicle $i$ to execute a local model update, and $T_{c_i}$ be the communication time needed to transmit the model updates to its neighboring vehicles. The total latency $L_i$ for vehicle $i$ is expressed as:
\begin{equation}
L_i = T_{p_i} + T_{c_i}.
\end{equation}
Considering the blockchain and FL framework, the processing time $T_{p_i}$ depends on the computational resources allocated to vehicle $i$, and the communication time $T_{c_i}$ depends on the allocated bandwidth, channel gain, and transmission power. Let $r_i$ denote the computational resource allocated to vehicle $i$, $g_{ij}$ represent the channel gain between vehicles $i$ and $j$, and $P_{ij}$ represent the transmission power from vehicle $i$ to vehicle $j$. The communication time is estimated as:
\begin{equation}
T_{c_i} = \frac{D_i}{B_i \log_2 \left(1 + \frac{P_{ij} g_{ij}}{N_0}\right)},
\end{equation}
where $D_i$ is the amount of data that needs to be transmitted from vehicle $i$, and $N_0$ is the noise power.

The objective is to minimize the total latency while maximizing the resource utilization. The optimization problem is formulated as:
\begin{align}
\text{minimize: } & \sum_{i=1}^N L_i \nonumber \\
\text{subject to:} \nonumber \\
& 0 \leq r_i \leq r_{\text{max},i}, \forall i \in \{1, \dots, N\} \label{eq:resource_constraint1} \\
& 0 \leq P_{ij} \leq P_{\text{max},ij}, \forall i, j \in \{1, \dots, N\} \label{eq:resource_constraint2} \\
& 0 \leq B_i \leq B_{\text{max},i}, \forall i \in \{1, \dots, N\} \label{eq:resource_constraint3} \\
& T_{p_i} \geq \frac{1}{r_i} f_i(X_i), \forall i \in \{1, \dots, N\} \label{eq:resource_constraint4}
\end{align}
where $r_{\text{max},i}$ is the maximum computational resource available for vehicle $i$, $P_{\text{max},ij}$ is the maximum transmission power from vehicle $i$ to vehicle $j$, $B_{\text{max},i}$ is the maximum available bandwidth for vehicle $i$, and $f_i(X_i)$ represents the workload of the local model update for vehicle $i$ based on its data $X_i$. This is a nonlinear programming problem, and it can be solved using various optimization techniques, such as gradient-based methods, evolutionary algorithms, or convex optimization techniques. To incorporate the parameters from the blockchain and FL model for V2V communication, we can adjust the constraints and objective function as follows:
\begin{itemize}
    \item Update the communication time $T_{c_i}$ to account for the consensus communication between vehicles during FL:
\begin{equation}
T_{c_i} = \frac{D_i}{B_i \log_2 \left(1 + \frac{P_{ij} g_{ij}}{N_0}\right)} + \sum_{k=1}^{K} \frac{D_{ik}}{B_i \log_2 \left(1 + \frac{P_{ij} g_{ij}}{N_0}\right)}
\end{equation}
where $K$ is the number of consensus rounds during FL, and $D_{ik}$ is the amount of data exchanged between vehicles $i$ and $j$ in round $k$.
    \item Introduce a constraint for the blockchain-based consensus to ensure the security and reliability of the FL:
\begin{equation}
\label{eq:resource_constraint5}
T_{\text{block}} \geq T_{\text{min}}, \forall i \in {1, \dots, N},
\end{equation}

where $T_{\text{block}}$ is the time required for the blockchain consensus process, and $T_{\text{min}}$ is a predefined minimum time to ensure the security of the blockchain-based consensus.

\item Update the objective function to include the impact of the blockchain consensus latency on the total latency:
\begin{equation}
\text{minimize: } \sum_{i=1}^N \left(L_i + T_{\text{block}}\right).
\end{equation}
\end{itemize}
The updated optimization problem can be solved using the same optimization techniques mentioned earlier. By solving this problem, we can obtain the optimal allocation of computational resources, transmission power, and bandwidth for each vehicle in the THz 6G network, which minimizes the total latency while maximizing the resource utilization for the blockchain-assisted FL framework in V2V communication.

\subsection{Integration \& Optimization}
To integrate the three complex mathematical models for blockchain, FL, and latency and resource allocation in a real-world V2V communication application, we propose a holistic optimization framework that combines these models to minimize latency, maximize resource utilization, and ensure secure and reliable communication in the 6G network.

The integrated optimization framework aims to minimize the overall latency while maximizing resource utilization and ensuring secure and reliable communication. The objective function combines the latency components from the FL model and the resource allocation model and includes the impact of the blockchain consensus latency.

Let $L_i^{\text{FL}}$ be the latency for vehicle $i$ in the FL framework, and $L_i^{\text{RA}}$ be the latency for vehicle $i$ in the resource allocation model. The total latency $L_i^{\text{total}}$ for vehicle $i$ is expressed as:
\begin{equation}
L_i^{\text{total}} = L_i^{\text{FL}} + L_i^{\text{RA}} + T_{\text{block}}.
\end{equation}

The integrated optimization problem is formulated as:
\begin{align}
\text{minimize: } & \sum_{i=1}^N L_i^{\text{total}} \nonumber \\
\text{subject to:} & \\
& \text{constraints from the blockchain model}  \nonumber \\
& \text{constraints from the FL model}  \nonumber \\
& \text{constraints from the resource allocation model} \nonumber
\end{align}

To solve this integrated optimization problem, we propose a two-stage optimization approach based on novel algorithms known as MOPSO and BACO.

\textbf{Stage-1: Multi-Objective Particle Swarm Optimization (MOPSO)}: 
To further detail the FL and Resource Allocation Optimization stage using evolutionary algorithm or state-of-the-art techniques, we leveraged Multi-Objective Particle Swarm Optimization (MOPSO).

Evolutionary algorithms are particularly suitable for multi-objective optimization problems since they can generate a diverse set of solutions, known as the Pareto front, representing trade-offs among conflicting objectives.

MOPSO is  multi-objective optimization Algorithm-\ref{one} which uses the concept of swarm intelligence to explore the search space for optimal solutions.
\begin{equation}
v_{i}^{t+1} = w v_{i}^{t} + c_{1} r_{1} (pbest_{i} - x_{i}^{t}) + c_{2} r_{2(gbest - x_{i}^{t})}
\end{equation}

\begin{equation}
x_{i}^{t+1} = x_{i}^{t} + v_{i}^{t+1}
\end{equation}
where $v_{i}^{t}$ and $x_{i}^{t}$ are the velocity and position of particle $i$ at time $t$, $w$ is the inertia weight, $c_{1}$ and $c_{2}$ are the cognitive and social learning factors, $r_{1}$ and $r_{2}$ are random numbers between 0 and 1, $pbest_{i}$ is the personal best position of particle $i$, and $gbest$ is the global best position.



\begin{algorithm}
\caption{Proposed MOPSO}
\label{one}
\begin{algorithmic}[1]
\State \textbf{Input:} $N_s$ (number of particles), $w$ (inertia weight), $c_1, c_2$ (cognitive and social learning factors), $MaxIter$ (maximum iterations).
\State (i)   Initialize the swarm with $N_s$ particles: $\mathcal{P} = \{P_1, P_2, \dots, P_{N_s}\}$.
\State (ii) Initialize the external archive $\mathcal{A} = \emptyset$
\For{$t = 1$ to $MaxIter$}
    \For{each particle $P_i \in \mathcal{P}$}
        \State (iii) Evaluate the objective function: $L_i^{\text{total}} = L_i^{\text{FL}} + L_i^{\text{RA}} + T_{\text{block}}$.
        \State (iv) Update the personal best position $pbest_i$.
    \EndFor
    \State (v) Compare each particle $P_i \in \mathcal{P}$: with solutions in $\mathcal{A}$ and update $\mathcal{A}$ as follows: 
        \State \quad \quad If $P_i$ is non-dominated, add it to $\mathcal{A}$. 
        \State \quad \quad If $P_i$ dominates any solutions in $\mathcal{A}$, remove the dominated solutions from $\mathcal{A}$.
    \State (vi) Maintain diversity in $\mathcal{A}$ using crowding distance or similar measure.
    \State (vii) Select the global best position $gbest$ from $\mathcal{A}$
    \For{each particle $P_i \in \mathcal{P}$}
        \State (viii) Update the velocity $v_i^{t+1}$ as:
            \begin{equation*}
            v_i^{t+1} = w v_i^t + c_1 r_1 (pbest_i - x_i^t) + c_2 r_2 (gbest - x_i^t)
            \end{equation*}
        \State Update the position $x_i^{t+1}$ as:
            \begin{equation*}
            x_i^{t+1} = x_i^t + v_i^{t+1}
            \end{equation*}
    \EndFor
\EndFor
\State \textbf{Output:} The Pareto-optimal solutions in the external archive $\mathcal{A}$, including their positions $x_i$ and velocities $v_i$.
\end{algorithmic}
\end{algorithm}

The MOPSO algorithm initializes particles and an archive. During each iteration, it evaluates objectives, updates personal best positions, and compares particles with archive solutions. Non-dominated particles are added, while dominated ones are removed. Diversity is maintained using crowding distance, and the global best position is selected. Velocities and positions are updated, and after a specified number of iterations (MaxIter), the Pareto-optimal solutions in the archive represent trade-offs for the integrated optimization problem.

\textbf{Stage-2: Blockchain-Assisted Consensus Optimization (BACO)}: In this stage, we optimize the blockchain-based consensus process considering the constraints from the blockchain model. We apply multi-agent reinforcement learning to develop a distributed and adaptive consensus algorithm that dynamically adjusts to the network conditions, while considering the latency, security, and reliability requirements.

The state space $\mathcal{S}$ contains local model updates, resource allocation parameters, and network conditions for consensus. The action space $\mathcal{A}$ includes actions like proposing blocks, validating transactions, and voting. Reward function $R: \mathcal{S} \times \mathcal{A} \rightarrow \mathbb{R}$ balances latency, security, and reliability, rewarding successful actions and penalizing invalid ones or high latency. The Deep Q-networks (DQN) train agents distributedly, each learning a Q-function and updating its policy to maximize rewards. Coordination mechanisms, such as communication protocols or shared models, avoid conflicts among agents. The BACO algorithm, leveraging multi-agent reinforcement learning, enables efficient and secure consensus while minimizing latency and maximizing reliability in V2V communications.



\begin{algorithm}
\caption{Proposed BACO using DQN}
\label{two}
\begin{algorithmic}[1]
\State \textbf{Input:} State space $\mathcal{S}$, action space $\mathcal{A}$, reward function $R$, number of episodes $E$, number of timesteps $T$, set of vehicles $\mathcal{N}$. 
\State \textbf{Initialization:}
\begin{itemize}
\item Initialize the state space $\mathcal{S}$, action space $\mathcal{A}$, and reward function $R$.
\end{itemize}
For each vehicle $i \in \mathcal{N}$:
\begin{itemize}
\item Initialize Q-network $Q_i$ with random weights $\theta_i$.
\item Initialize target Q-network $\tilde{Q}_i$ with weights $\tilde{\theta}i = \theta_i$.
\item Initialize replay buffer $\mathcal{D}i$.
\end{itemize}
\For{episode $e = 1, 2, \dots, E$}
\State Observe initial state $s_0$ and broadcast it to all vehicles.
\For{timestep $t = 1, 2, \dots, T$}
\For{each vehicle $i \in \mathcal{N}$ (in parallel)}
\State Select action $a_i^t$ using an $\epsilon$-greedy policy based on $Q_i(s_{t-1}, \cdot; \theta_i)$.
\State Execute action $a_i^t$ and observe local reward $r_i^t$ and next state $s_i^t$.
\State Store transition $(s_{t-1}, a_i^t, r_i^t, s_i^t)$ in replay buffer $\mathcal{D}_i$.
\State Sample a mini-batch of transitions $(s_j, a_j, r_j, s'_j)$ from $\mathcal{D}_i$.
\State Compute target values $y_j = r_j + \gamma \max_{a'} \tilde{Q}_i(s'_j, a'; \tilde{\theta}_i)$.
\State Update Q-network weights $\theta_i$ by minimizing the loss: $\mathcal{L}(\theta_i) = \frac{1}{|\mathcal{D}_i|} \sum_j (y_j - Q_i(s_j, a_j; \theta_i))^2$.
\State Update target Q-network weights: $\tilde{\theta}_i \leftarrow \tau \theta_i + (1 - \tau) \tilde{\theta}_i$, where $\tau$ is a small constant.
\EndFor
\State Update global state $s_t$ by aggregating local states $s_i^t$ from all vehicles.
\EndFor
\EndFor
\State \textbf{Output:} Trained Q-networks for each vehicle $i \in \mathcal{N}$. 
\end{algorithmic}
\end{algorithm}

Algorithm \ref{two}, BACO using DQN, initializes the state space, action space, reward function, and Q-networks for each vehicle. It allows each vehicle (agent) to learn an optimal policy for participating in the blockchain consensus process while considering the latency, security, and reliability trade-offs. 
The algorithm runs through episodes and timesteps, where vehicles select and execute actions, observe rewards and next states, store transitions, and update Q-network. Target Q-network weights are also updated and the global state is aggregated from local states. The output consists of trained Q-networks for each vehicle for their decision-making policies.


\section{Experiment Results and Analysis} \label{S}
This section presents the performance evaluation of our proposed framework regarding the system efficiency, reliability, security, and stability metrics. We also include the comparison results with other recent studies. In particular, we formulated three candidate for comparisons as \textbf{State of the art technique-1} based on \cite{9798266}, \textbf{State of the art technique-2} based on \cite{9990577}, and \textbf{Dynamic CU-DU} selection referred to \cite{mollahasani2021dynamic}. Our method is referred to as the \textbf{Proposed System Model}.

\begin{table}[h]
\centering
\caption{Simulation Parameters}
\label{tab:sim_params}
\begin{tabular}{l|c}
\hline
\textbf{Parameter} & \textbf{Value} \\ \hline
Number of Vehicular Nodes & 100 \\ \hline
Communication Range & 500 meters \\ \hline
Packet Size & 1500 bytes \\ \hline
Simulation Time & 3560 seconds \\ \hline
Number of Simulation Runs & 33 \\ \hline
Data Collection Interval & 1 second \\ \hline
Blockchain Block Size & 1 MB \\ \hline
Blockchain Block Interval & 10 seconds \\ \hline
Number of FL Rounds & 9 \\ \hline
Learning Rate & 0.01 \\ \hline
Batch Size & 32 \\ \hline
Terahertz Carrier Frequency & 300 GHz \\ \hline
Terahertz Bandwidth & 10 GHz \\ \hline
\end{tabular}
\vspace{-3mm}
\end{table}

\subsection{Efficiency Metrics}
\begin{figure}[htbp]
  \centering
  
  \begin{subfigure}[b]{0.49\linewidth}
    \includegraphics[width=\linewidth]{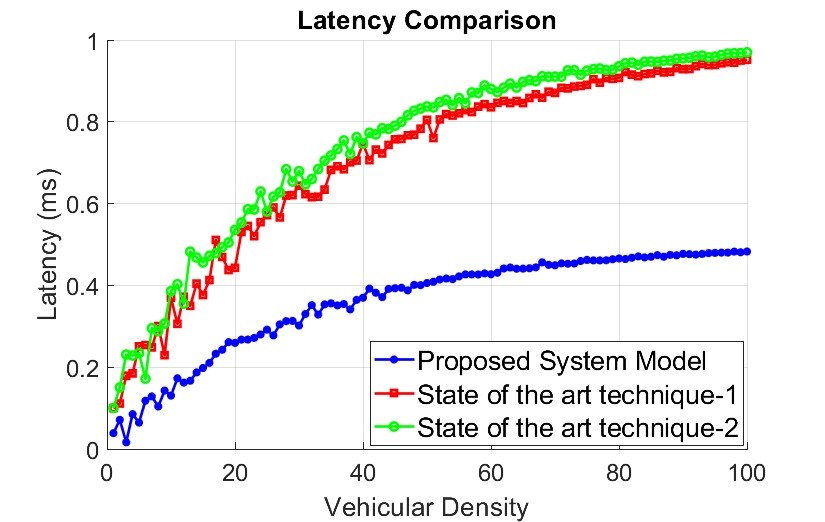}
    \caption{{Latency}}
    \label{fig:Latency}
  \end{subfigure}
  \hfill
  \begin{subfigure}[b]{0.49\linewidth}
    \includegraphics[width=\linewidth]{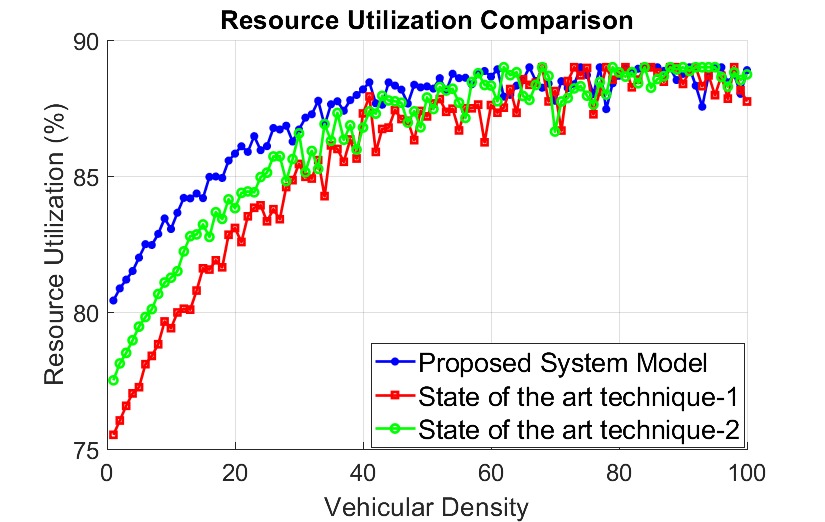}
    \caption{{Resource Utilization}}
    \label{fig:RUP}
  \end{subfigure}

  \caption{System Latency and Resource Utilization results}
  \label{fig:efficiency}
  \vspace{-6mm}
\end{figure}

The system efficiency is evaluated based on the latency and resource utilization metrics. As depicted in Figure \ref{fig:efficiency}, the proposed V2V communication system outperforms other techniques in efficiency metrics. The model showcases a remarkable latency reduction, clocking in at less than 0.1 millisecond - a significant improvement over conventional methods. Furthermore, the system's resource utilization competes effectively with other models, signaling adept resource management. It's worth noting that elevated resource utilization values could denote intricate traffic congestion, a factor intricately tied to the packet delivery ratio (PDR).

\subsection{Reliability Metrics}
\begin{figure}[htbp]
  \centering
  
  \begin{subfigure}[b]{0.49\linewidth}
    \includegraphics[width=\linewidth]{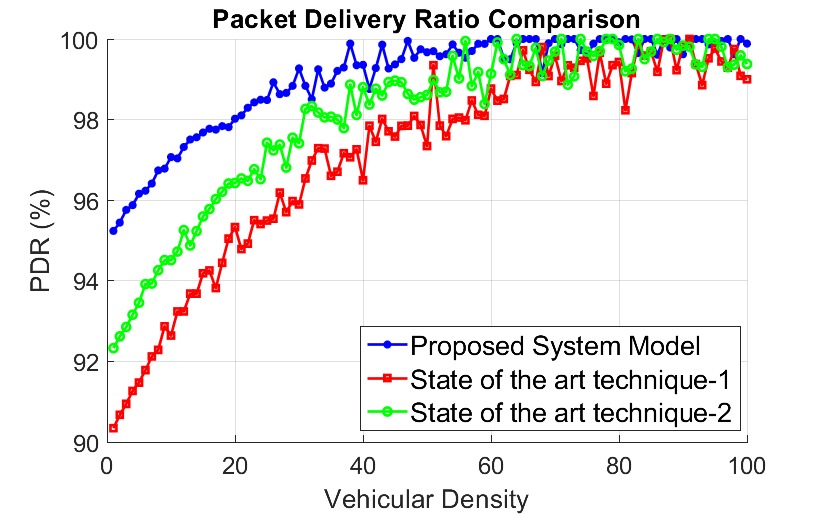}
    \caption{{PDR}}
    \label{PDR}
  \end{subfigure}
  \hfill
  \begin{subfigure}[b]{0.49\linewidth}
    \includegraphics[width=\linewidth]{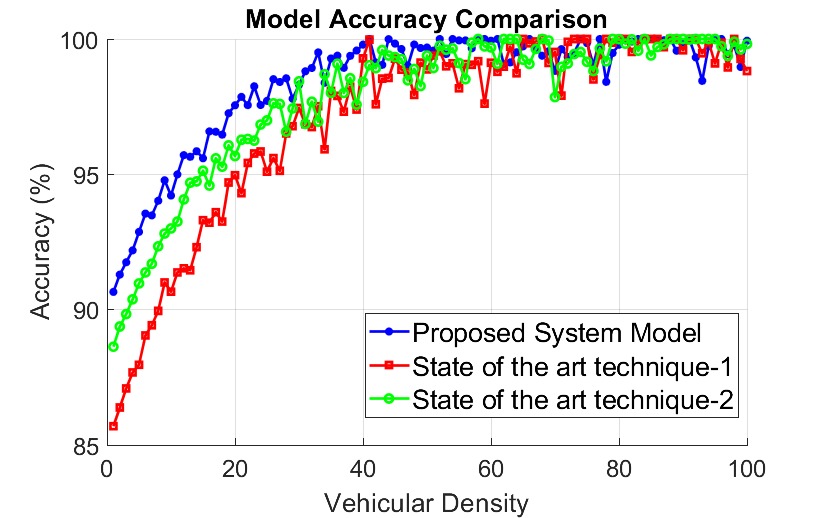}
    \caption{{Model Accuracy}}
    \label{MA}
  \end{subfigure}

  \caption{Measure of PDR and Model Accuracy}
  \label{fig:Reliabi}
   \vspace{-3mm}
\end{figure}

To evaluate the system reliability, we consider the measurement of Packet Delivery Ratio (PDR) and federated model accuracy regarding vehicular density (see Figure \ref{fig:Reliabi}). We also evaluate the Mean Time to Failure (MTTF) (see Figure \ref{3a}). The results reveal that our proposed system model enhanced reliability better than the other techniques. Even amid increased vehicular densities, the PDR for our system maintained a consistently high, unlike other techniques which exhibited a discernible dip. Furthermore, our system demonstrated superior federated model accuracy ($>94\%$), suggesting a more efficient learning mechanism from the local data of connected vehicles. The proposed system excels with an impressive MTTF of 6400 hours, surpassing the 6200-hour benchmark, showcasing superior failure minimization and system dependability. 

\subsection{Assessing System Security and Stability}
\begin{figure}[htbp]
  \centering
  
  \begin{subfigure}[b]{0.49\linewidth}
    \includegraphics[width=\linewidth]{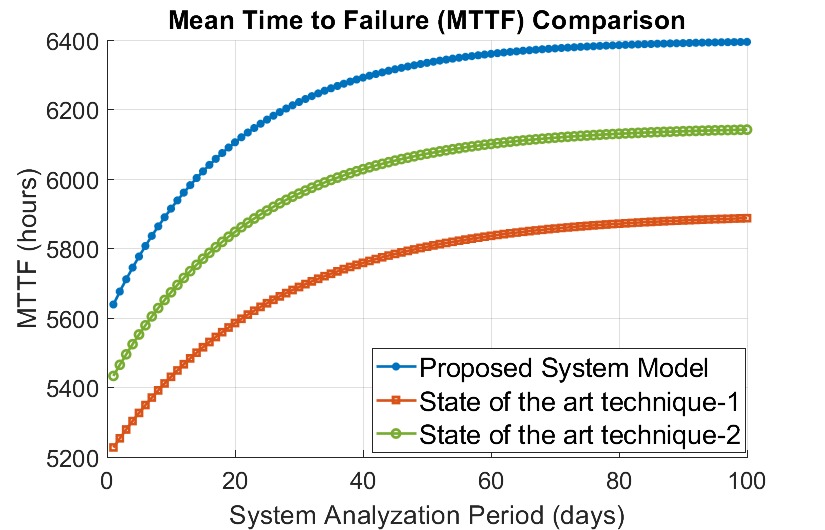}
    \caption{MTTF}
    \label{3a}
  \end{subfigure}
  \hfill
  \begin{subfigure}[b]{0.49\linewidth}
    \includegraphics[width=\linewidth]{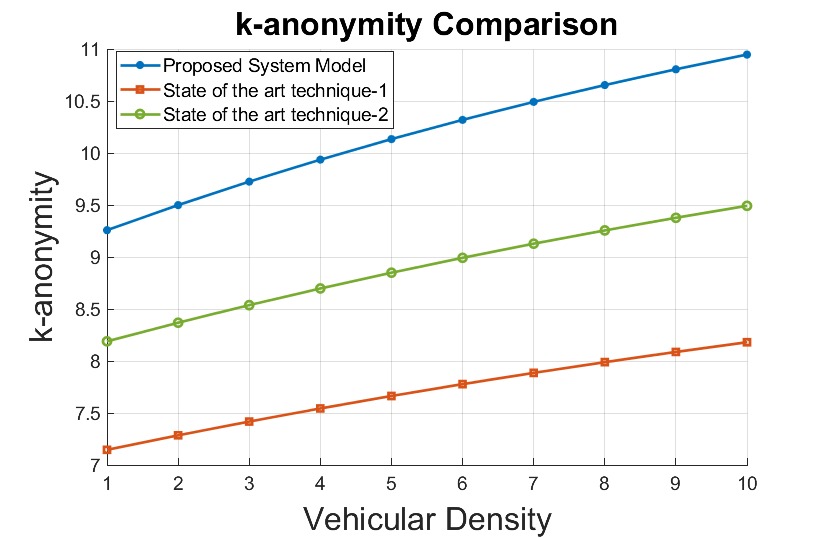}
    \caption{{k-anonymity}}
    \label{3b}
  \end{subfigure}

  \caption{ Mean Time to Failure and  k-anonymity measurements}
  \label{fig:security}
  \vspace{-3.5mm}
\end{figure}

We assess the system security through k-anonymity metric. K-anonymity is incorporated by anonymizing the vehicle data to ensure that each record cannot be distinguished from at least k-1 other records. The findings are depicted in Figure \ref{3b}. 
The proposed system can achieve a k-anonymity factor exceeding 9, which ensures high privacy preservation. 

\begin{figure}[htbp]
  \centering
  
  \begin{subfigure}[b]{0.49\linewidth}
    \includegraphics[width=\linewidth]{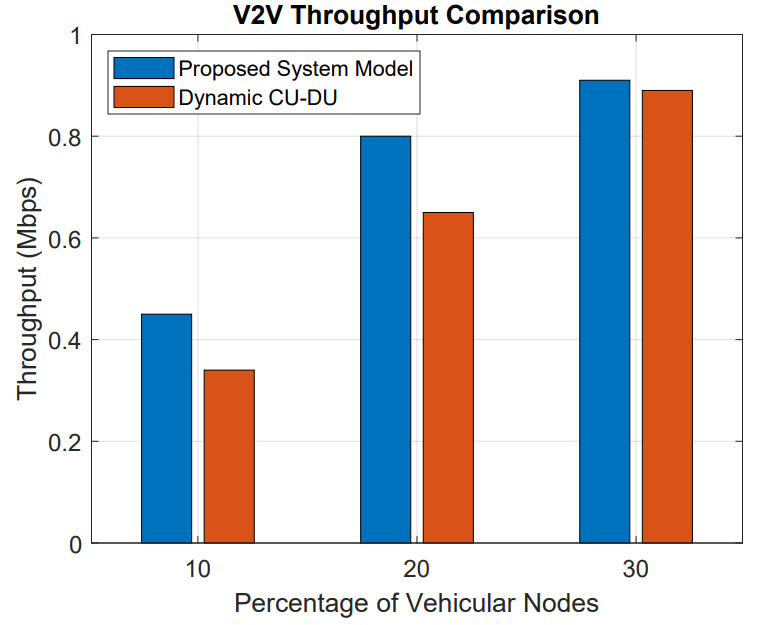}
    \caption{{V2V Throughput}}
    \label{4a}
  \end{subfigure}
  \hfill
  \begin{subfigure}[b]{0.49\linewidth}
    \includegraphics[width=\linewidth]{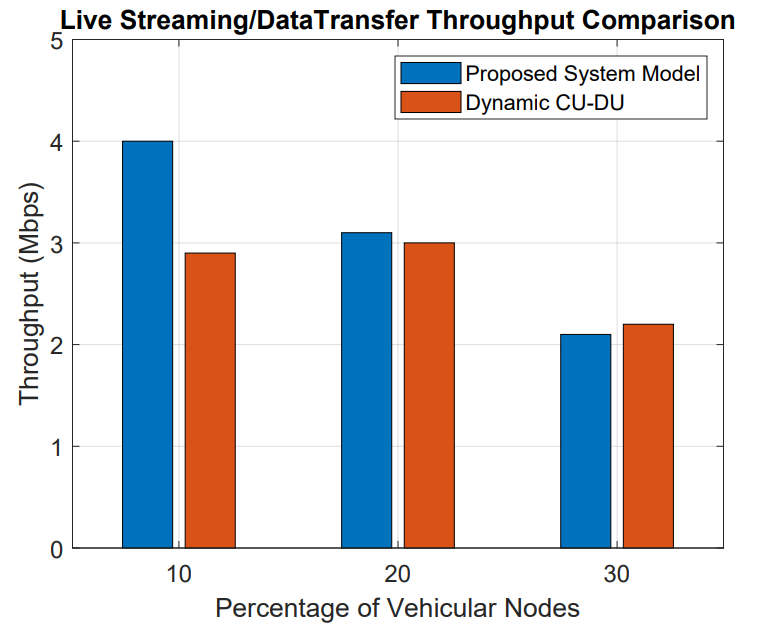}
    \caption{Streamlining Throughput}
    \label{4b}
  \end{subfigure}

  \caption{Measure of System Stability}
  \label{4}
  \vspace{-6mm}
\end{figure}

Furthermore, an analysis of the system's throughput is conducted to gauge the robustness and stability of the proposed model. Throughput is evaluated in two dimensions and juxtaposed with the cutting-edge Dynamic {CU-DU} selection for resource allocation in {O-RAN} using actor-critic learning. As illustrated in Figure \ref{4a}, our proposed model triumphs with a superior overall throughput of 0.91, surpassing the current method's 0.89 at maximum vehicular density. We also scrutinize the live streaming throughput among connected vehicles. Our model exhibits competitive performance across a multitude of scenarios, as vividly represented in Figure \ref{4b}.

\section{Conclusion} \label{C}

This research presents a sophisticated V2V communication system for terahertz 6G networks, featuring an integration of blockchain, FL, and resource optimization. The proposed MOPSO and BACO strategies significantly advance this system, exhibiting enhancements in latency, resource utilization, consensus security, and data privacy beyond traditional and 5G technologies. Further incorporation of advanced methodologies such as game theory and multi-agent reinforcement learning could potentially deepen the impact of this system, setting the stage for transformative advancements in 6G V2V networks.

\section*{Acknowledgment}
The research work of Tahir, Tiang, and Azwan is Funded
by the Ministry of Higher Education, Malaysia, under
the number MMUE/190229 (FRGS/1/2020/TK02/MMU/03/1) and MMUE/210058 (FRGS/1/2021/TK02/MMU/02/1). All
authors have no conflicts of interest to declare

\bibliographystyle{IEEEtran}
\bibliography{References.bib}
\end{document}